\documentclass{llncs}

\usepackage{graphicx}
\usepackage{url}
\usepackage{amsmath}
\usepackage{amssymb}
\usepackage{amsfonts}
\newtheorem{katei}{Assumption}

\begin{document}
\title{o-glasses: Visualizing x86 Code from Binary Using a 1d-CNN}
\author{Yuhei Otsubo\inst{1}\inst{2}
\and Akira Otsuka\inst{2}
\and Mamoru Mimura\inst{3}\inst{2}
\and Takeshi Sakaki\inst{4}
\and Atsuhiro Goto\inst{2}}
\institute{National Police Agency, Tokyo, Japan
\and Institute of Information Security, Kanagawa, Japan\\
\email{dgs157101@iisec.ac.jp}
\and National Defense Academy, Kanagawa, Japan
\and The University of Tokyo, Tokyo, Japan}

\maketitle

\begin{abstract}
Malicious document files used in targeted attacks often contain a small program called shellcode.
It is often hard to prepare a runnable environment for dynamic analysis of these document files because they exploit specific vulnerabilities.
In these cases, it is necessary to identify the position of the shellcode in each document file to analyze it.
If the exploit code uses executable scripts such as JavaScript and Flash, it is not so hard to locate the shellcode.
On the other hand, it is sometimes almost impossible to locate the shellcode when it does not contain any JavaScript or Flash but consists of native x86 code only.

Binary fragment classification is often applied to visualize the location of regions of interest, and shellcode must contain at least a small fragment of x86 native code even if most of it is obfuscated, such as, a decoder for the obfuscated body of the shellcode.
In this paper, we propose a novel method, o-glasses, to visualize the shellcode by recognizing the x86 native code using a specially designed one-dimensional convolutional neural network (1d-CNN).
The fragment size needs to be as small as the minimum size of the x86 native code in the whole shellcode.
Our results show that a 16-instruction-sequence (approximately 48 bytes on average) is sufficient for the code fragment visualization.
Our method, o-glasses (1d-CNN), outperforms other methods in that it recognizes x86 native code with a surprisingly high F-measure rate (about 99.95\%).
\end{abstract}

Binary fragment classification, shellcode, visualization, CNN, MLP 

\section{Introduction}
In recent years, targeted attacks have become a major threat.
In a targeted email attack, an email contains a request to open an attached file or click on a hyperlink in the email body. If the recipient does so, then some malware is launched.
Most such malware is newly crafted, unknown malware, and is thus often hard for antivirus scanners to catch. In particular, malicious document files used in targeted email attacks often contain an executable file embedded within a decoy document file:
over 60\% of the attached files in targeted email attacks occurring in 2014 were reported to be document files~\cite{trend-annual2014}.

\begin{figure}[tb]
\begin{center}
  \includegraphics[width=8cm,clip]{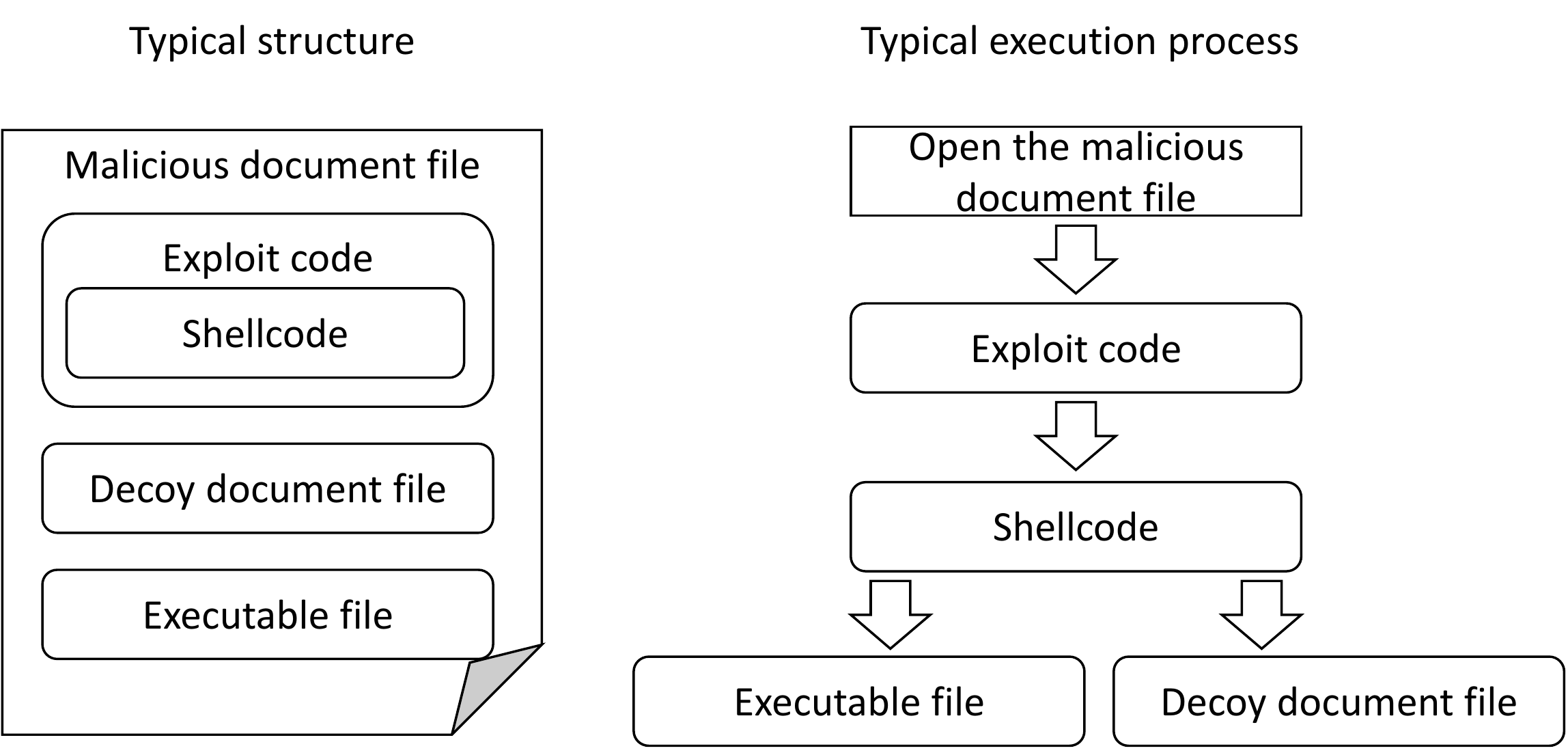}
\caption{Typical structure and execution process of a malicious document}
\label{fig:mal_doc}
\end{center}
\end{figure}
The left-hand side of Fig.~\ref{fig:mal_doc} shows a typical structure for a malicious document file.
The malicious document file consists of four parts: exploit code, shellcode, an executable file, and a decoy document file.
Exploit code is a program designed to exploit a document processor vulnerability.
The exploit code is executed when the malicious document file is opened, leading to execution of the shellcode.
Shellcode is a program designed to create an executable file and a decoy document from the remainder of the file and to launch the executable.
Then, the PC that opened the malicious document file becomes controllable by attackers.
The right-hand side of Fig.~\ref{fig:mal_doc} shows a typical execution process of a malicious document file.

To reach attackers' information, we should not only detect the malware but also figure out the features of the malware in detail.
Here, we face several problems.
First, we should prepare a runnable condition for the malware in order to conduct dynamic analysis.
When the target file is a malicious document exploiting specific vulnerabilities, it is often difficult to prepare the activatable environment (OS versions, browsing software, language, patches, and so on) because the conditions are complicatedly intertwined.
Therefore, we are often forced to conduct static analysis.
When the target file is an executable file, it is easy to find the entry point for analysis.
However, when the target file is a document file, it is not so easy to find the entry point.
In this case, we focus on the shellcode executed after exploit code.
When the malware uses JavaScript or Flash, we can figure out the location of the shellcode quickly.
However, exploit code uses not only JavaScript and Flash but also font and image files, for example, a TIFF image (CVE-2017-5133~\cite{cve-2017-5133_1,cve-2017-5133_2}, CVE-2004-1308~\cite{cve-2004-1308}), a jpeg2000 image (CVE-2016-8332~\cite{cve-2016-8332_1,cve-2016-8332_2}), and a TrueType font (CVE-2011-3402~\cite{cve-2011-3402}).
When searching for shellcode, it is important to consider various types of exploit code.
Thus, our target is a class of malicious document files that contain x86 native code hidden somewhere in them.

\begin{figure}[tb]
\begin{center}
  \includegraphics[width=8cm,clip]{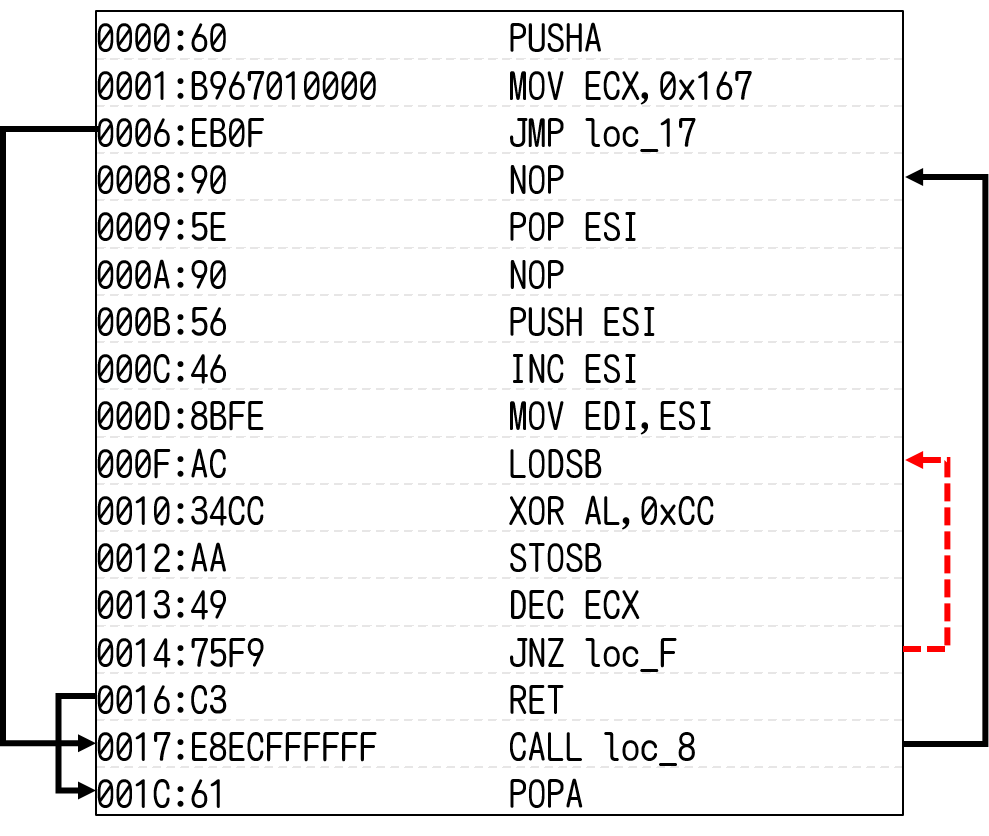}
\caption{
An example of a small decoder with a 29-byte sequence.
It contains only 17 opcodes, and it decodes the body of the shellcode with 1-byte-key xor encoding.
(``0xCC'' in this example)
}
\label{fig:shellcode}
\end{center}
\end{figure}

Although attackers tend to use obfuscation to protect their code, shellcode must contain at least a small fragment of x86 native code, such as a decoder. 
Fig.~\ref{fig:shellcode} shows an example of a small decoder containing 17 opcodes in only a 29-byte sequence.
This code was obtained from a malicious document file with a size of more than 100kB used in a real attack.
Our challenge, therefore, is finding a small amount of code like that shown in Fig.~\ref{fig:shellcode} in often large document files.
To do this, we introduce a novel method, called o-glasses, to visualize the shellcode by recognizing the x86 native code using a specially designed one-dimensional convolutional neural network (1d-CNN).

In summary, the main contributions of our approach are as follows:

\subsection{Easily Collectible Training Datasets}
One of the most significant problems in using machine learning is how to prepare the training dataset.
Even an excellent model cannot demonstrate its performance without large samples.
However, studies of malware using machine learning sometimes struggle to collect samples because they need examples of malware for training.
In contrast, our approach does not need malware for the training dataset.
Thus, samples for learning are easily available for anyone.

\subsection{High Recognition Rate for x86 Code}
Conventional signature-based malware detectors do not work when an unknown code is embedded.
On the other hand, program code is not supposed to be in document files.
So, extracting shellcode from malicious document files becomes a reality if we can separate program code precisely from normal byte sequences in document files.
The solution provided in this paper is based on the assumption that shellcode and general program code have similar distributions of code.

\subsection{Visual Analysis for Supporting Analysts}
Visualizing a binary as an image helps to quickly get an overview of the file.
While some experienced analysts can deduce the location of the embedded program code from a grayscale image converted from the binary file, even unexperienced analysts can achieve similar results using our proposed methods.

\section{Preliminaries}
The proposed solution lies in the static analysis of files.
We do not take into account the file structure.
The only thing of importance for us is whether a file fragment is a piece of x86 code.

\subsection{x86 Architecture}
The x86 and x86-64 architectures are probably the most widely used CISC (Complex Instruction-Set Computing) architectures~\cite{IntelManual}.
Their instruction sets are rich and complex, and most importantly they support instructions of varying length.
Instruction lengths range from just one byte (i.e., instructions comprising just a one-byte opcode) to 15 bytes.

\subsection{Assumptions}
We made the following assumptions.
\begin{katei}
The distribution of the byte sequence from x86 code is dissimilar to that from document files.
\end{katei}
\begin{katei}
The distribution of shellcode is the same as that of common x86 code.
\end{katei}
In other words, we expect to be able to detect shellcode by detecting any x86 code.

We next describe Shannon entropy, conventional visualization methods, and the deep learning models (multi-layer perceptron [MLP] and CNN) used in the study.

\subsection{Shannon Entropy}
We calculate the information entropy of each file fragment using the Shannon entropy rate given by
\begin{equation}
\label{equ:shannon}
{H(X)= -\frac{1}{8}\sum_{i=0}^{255}P(X=i)\log_{2}P(X=i)
},
\end{equation}
where $X$ is a random variable over [0, 255].
The entropy rates are real numbers between 0 and 1, where 1 means the file fragment is uniformly random.

\subsection{Conventional Visualization Methods}
Visualizing a binary as an image is very helpful for getting a quick overview of the file.
In this section, we describe the three conventional visualization methods.

\subsubsection{Grayscale}
A technique for representing different files with grayscale images was introduced by Conti \emph{et al.}~\cite{conti2010visual} and was applied to automatic malware classification by Nataraj \emph{et al.}~\cite{nataraj2011malware}.

\subsubsection{Bit-image representation of a binary file}
Goto~\cite{Stirling} implemented the visualization of a binary file as a ``bit-image'' in a hex editor named ``Stirling'' in 1998.
In Stirling, a given binary is read as a vector of 8-bit unsigned integers and then organized into a two-dimensional array.
This can be visualized as a bit-image in four colors: 0x00 (null) in white, 0x01-0x1F (control characters) in light blue, 0x20-0x7F (ASCII) in red, and 0x80-0xFF in black.

\subsubsection{Structural entropy}
Document files contain data of various kinds: metadata, text, and packed data.
All of these file areas differ not only in size but also in the level of information entropy.
When a document file may be considered as a system of such elements, then we can use the term structural entropy for its characterization.
Sorokin~\cite{Sorokin2011} built entropy diagrams by using the sliding window method.
He selected 256 bytes for the window (block) size and 128 bytes for the window (block) shift.
In our experiment, we used the same block size but changed the block shift to 1 byte to provide more detail.
We calculate entropy level at each offset and visualize the structural entropy as a grayscale image.

\subsection{MLP}
A standard MLP neural network has a three-tier structure: the input layer, the hidden layers, and the output layer.
Every layer in an MLP consists of nodes fully connected with the nodes in the adjacent layer. 

\subsection{CNN}
\begin{figure}[tb]
\begin{center}
  \includegraphics[width=8cm,clip]{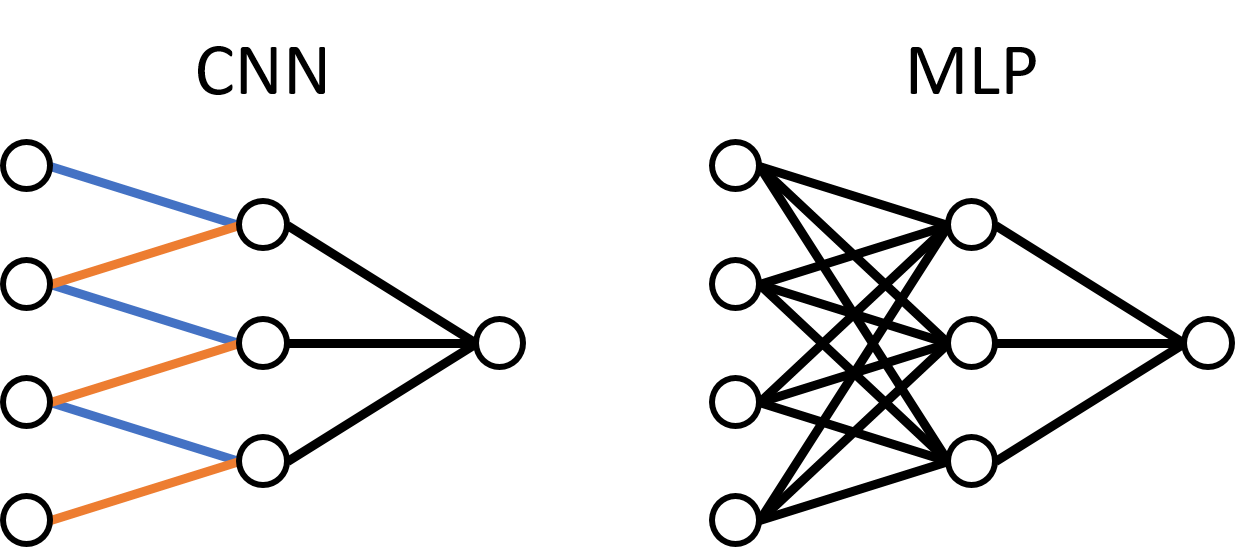}
\caption{Schematic diagrams of a CNN and an MLP}
\label{fig:cnn}
\end{center}
\end{figure}

In our method for recognizing x86 native code, we use a 1dCNN~\cite{doi:10.1162/neco.1989.1.4.541}.
In contrast to an MLP, a CNN has limited connections between each layer (see Fig.~\ref{fig:cnn}) and nodes in an intermediate layer receive only input from a localized part of the previous layer, which is called the receptive field.

Tools based on CNNs have now led to great results in a wide range of vision tasks~\cite{NIPS2012_4824}.
Generally, image data are continuous data.
So, when image data are input to a CNN, high object recognition power can be obtained by adjusting the CNN's local receptive field.
On the other hand, program code is classified as discrete data when viewed one byte at a time.
Therefore, when binary data directly converted into an image are input to a CNN, there is the possibility that the benefit of the local receptive field cannot be obtained.
On the other hand, program code is a sequence of instructions, which may reduce the variation, so the possibility of receiving the benefit of the CNN's local receptive field is not entirely ruled out.

Weight sharing is a mechanism in which all links to nodes of a local receptive field have the same weight.
In the case of Fig.~\ref{fig:cnn}, the three blue links have the same weight.
Similarly, the three red links have the same weight.
By using the local receptive field in this way, the result for some input data is the same as the result for shifted input data. This allows us to reflect all the data in intermediate layers despite the limited connectivity to the input.

\begin{figure}[tb]
\begin{center}
  \includegraphics[width=8cm,clip]{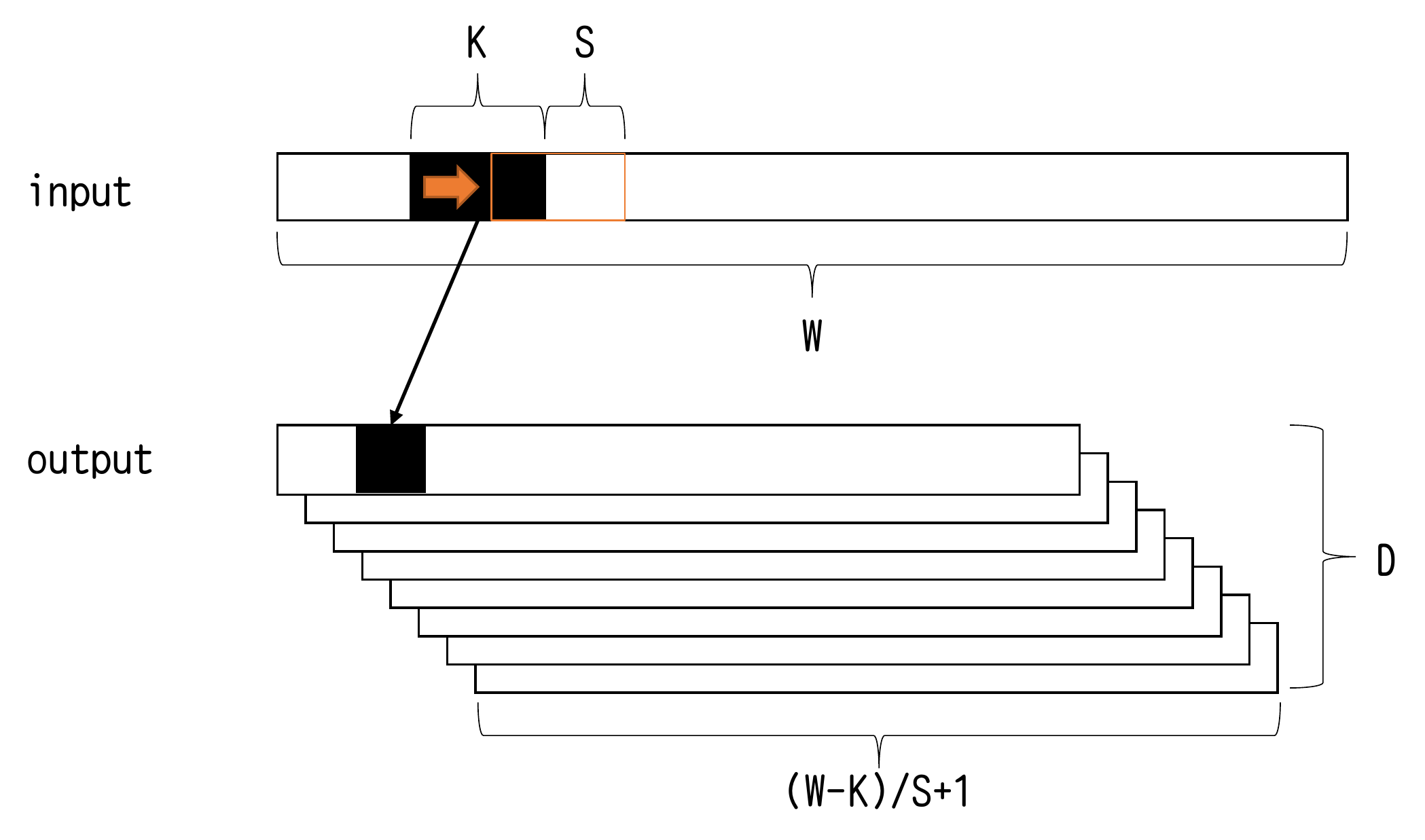}
\caption{Illustration of the one-dimensional convolutional architecture}
\label{fig:cnn_2}
\end{center}
\end{figure}

Several hyperparameters control the size of the output volume of the convolutional layer (Fig.~\ref{fig:cnn_2}):
the kernel field size, the depth, stride, and zero-padding.
We will ignore zero-padding because we do not use it.
The depth ($D$) of the output volume controls the number of neurons in a layer that connect to the same region of the input volume.
The stride ($S$) controls how depth columns around the spatial dimensions (width and height) are allocated.

The spatial size of the output volume can be computed as a function of the input volume $W$, the kernel field size of the convolutional layer neurons $K$, and the stride with which they are applied $S$.
The formula for calculating how many neurons ``fit'' in a given volume is given by $(W-K)/S+1$.

\section{Related Work}
Methods of analyzing malware can be divided into two types: static analysis and dynamic analysis.
We focus on static analysis as explained previously.

OfficeMalScanner~\cite{officemalscanner} (OMS) is an analysis tool for document files.
OMS scans entire files for generic shellcode patterns, an embedded signature of document files, or an embedded executable file.
Although this method incorporates fuzzy search, it is easy to avoid detection because the number of the search patterns is small.

MDScan~\cite{DBLP:conf/eurosec/TzermiasSPM11} is a standalone malicious document scanner. The tool analyzes PDF document files individually and detects malicious code.
The tool combines static analysis of the document format representation and dynamic analysis of the embedded script code.
The method focuses only on JavaScript in PDF files.
Hence, the method does not work well when the exploit code is not written in JavaScript.

There are several approaches to malware detection that use binary or grayscale images (binary texture analysis~\cite{nataraj2011comparative}, malware images~\cite{nataraj2011malware}, support vector machines~\cite{kancherla2013image} and visualization of binary files~\cite{han2013malware}).
These approaches are aimed toward the detection and the classification of malicious software based on image processing techniques.
Hence, they do not focus on finding a small amounts x86 code, such as shellcode, as we are doing here.

Binary fragment classification can visualize the location of regions of interest.
The fragment size needs as small as the size of shellcode to find it.
Xu \emph{et al.}~ \cite{xu2014file} treated a 1024-byte file fragment as a grayscale image and used an image classification method to classify file fragment.
They focused on file type classification for digital forensics.
It is difficult to make the fragment smaller because the texture of its grayscale image becomes harder to analyze.
Hence, it is difficult to find shellcode using this method.

\section{Training Data}
\label{sec:dataset}
\begin{figure}[tb]
\begin{center}
  \includegraphics[width=10cm,clip]{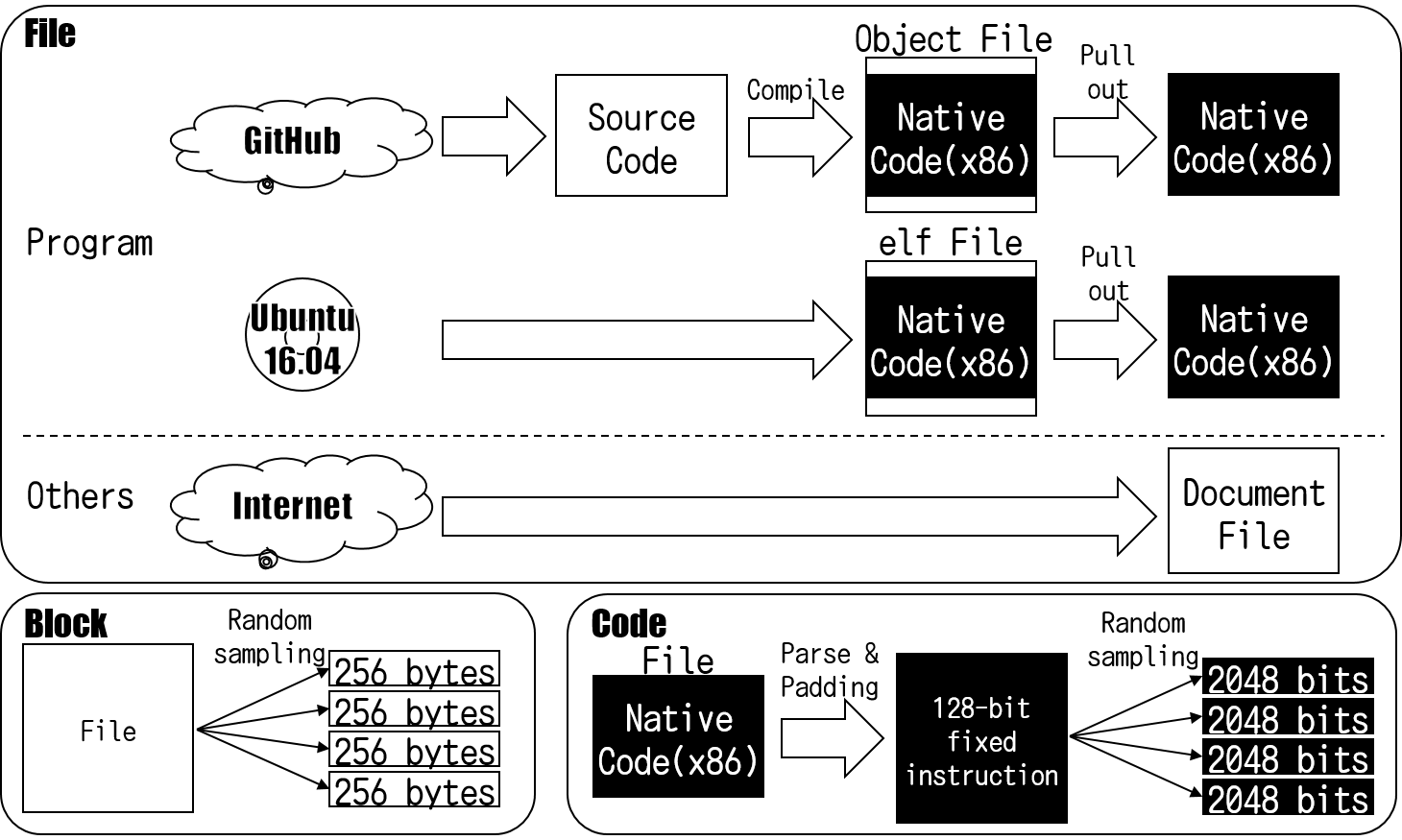}
\caption{Method for reproducing our datasets}
\label{fig:make_dataset}
\end{center}
\end{figure}

We prepared two categories of dataset for training,
both of which can be gathered easily.
One category is labelled ``Program'' and comprises various sets of x86 code taken from two sources: Github and Ubuntu 16.04.
The other category is called ``Others'' and consists of various document files and portions of data extracted from them.
The ``Others'' category contains ``CFB,''``OOXML,'' and ``PDF'' files.
CFB stands for compound file binary~\cite{cfb}, and it is used as a container like the FAT16 file system.
CFB is used in files with the extensions ``.doc,'' ``.xls,'' ``.ppt,'' ``.jtd'' (used by the ``Ichitaro'' Japanese word processor), ``.hwp'' (used by the ``Araea Han-geul'' Korean word processor), and so on.
OOXML stands for Office Open XML~\cite{ooxml}, which is a zip container in reality.
OOXML is used in ``.docx,'' ``.xlsx,'' and ``.pptx'' files.
PDF stands for portable document format~\cite{pdf}, which has the extension ``.pdf.'' For each category and source or file type, we constructed there types of dataset: the whole files, 256-byte blocks extracted from these files, and 2048-bit segments of code extracted from the files.
Table~\ref{tab:dataset} shows an outline of our datasets.
\begin{table}[tb]
\caption{Number of elements in each of our datasets} 
\label{tab:dataset}
\begin{center}
\begin{tabular}{cc|p{5em}p{5em}p{5em}}
\hline
\hline
&& \multicolumn{3}{|c}{Type of dataset}\\
\multicolumn{2}{c|}{Category \& Source} & \hfill File & \hfill Block & \hfill Code \\
\hline
Program & GitHub & \hfill 1,147 & \hfill 49,577 & \hfill 258,793\\
& Ubuntu & \hfill 295 & \hfill 72,103 & \hfill 404,427\\
& Total & \hfill 1,442 & \hfill 121,680 & \hfill 663,220\\
\hline
Others  &CFB & \hfill 27 & \hfill 31,779 & \hfill 223,904\\
& OOXML & \hfill 18 & \hfill 33,495 & \hfill 220,669\\
& PDF & \hfill 27 & \hfill 35,357 & \hfill 232,573\\
& Total & \hfill 72 & \hfill 100,631 & \hfill 677,146\\
\hline
\multicolumn{2}{c|}{Total} & \hfill 1,514 & \hfill 222,311 & \hfill 1,340,366\\
\hline
\end{tabular}
\end{center}
\end{table}
An outline of how to make our datasets is shown in Fig.~\ref{fig:make_dataset}.
The methods for making each of our types of dataset are as follows.

\subsubsection{File}
The following procedure is conducted for making the file dataset in the ``Program:GitHub'' category.
\begin{itemize}
 \item Gather various C/C++ source code files from GitHub~\cite{github}
 \item Compile these files into x86 object files by using gcc~\cite{gcc}
 \item Extract only the native code from these object files.
\end{itemize}

To make the file dataset in the ``Program:Ubuntu'' category, we extracted program code from the elf files in the ``/bin'' and ``/sbin'' directories of Ubuntu 16.04 using the header information.

Finally, to make the file datasets in the ``Others'' category, we used a search engine to gather various open-source document files.
Table~\ref{tab:keyword} shows the keywords used for this search.
\begin{table}[tb] 
\caption{The keyword list for each label} 
\label{tab:keyword}
\begin{center}
\begin{tabular}{cc|p{9em}}
\hline
\hline
&& keyword list\\
\hline
&CFB & ``test'',``.doc''\\
Others&OOXML & ``test'',``.docx''\\
&PDF& ``test'',``.pdf''\\
\hline
\end{tabular}
\end{center}
\end{table}
We downloaded document files from the beginning of the list of search results.
We then checked these download files using VirusTotal~\cite{virustotal}, and we removed suspicious files that were detected as malware.

\subsubsection{Block}
\begin{figure}[tb]
\begin{center}
  \includegraphics[width=8cm,clip]{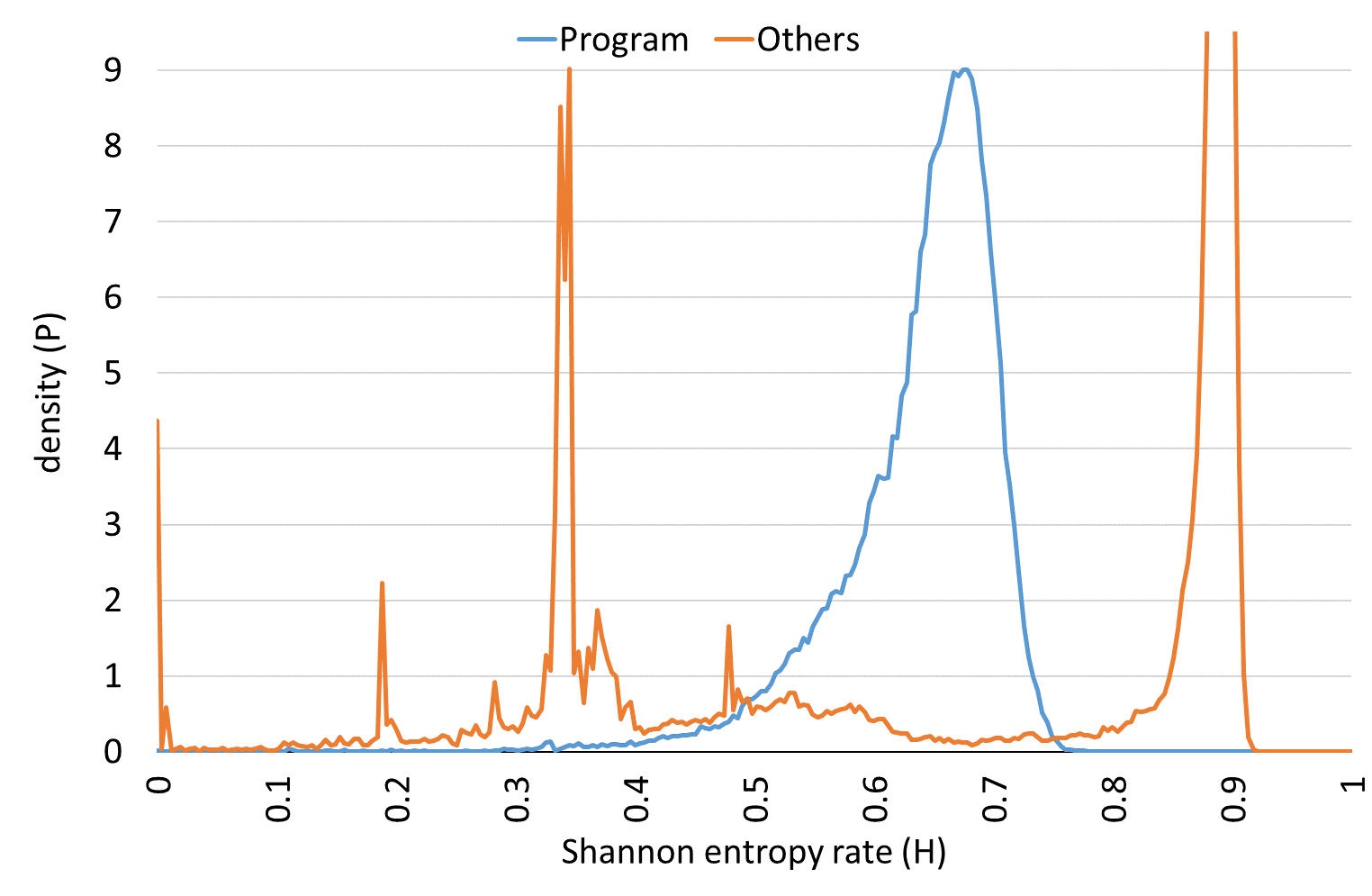}
\caption{Distributions of the Shannon entropy rate for blocks in the two categories of dataset}
\label{fig:entropy}
\end{center}
\end{figure}

``Block'' datasets are made by extracting 256-byte blocks by random sampling from every file in a ``File'' dataset.
We calculated Shannon entropy rates (Equation (\ref{equ:shannon})) for each block in the ``Block'' datasets.
The distributions of the entropy rates for blocks in the ``Program'' and ``Others'' categories are shown in Fig.~\ref{fig:entropy}.

\subsubsection{Code}
\begin{figure}[tb]
\begin{center}
  \includegraphics[width=8cm,clip]{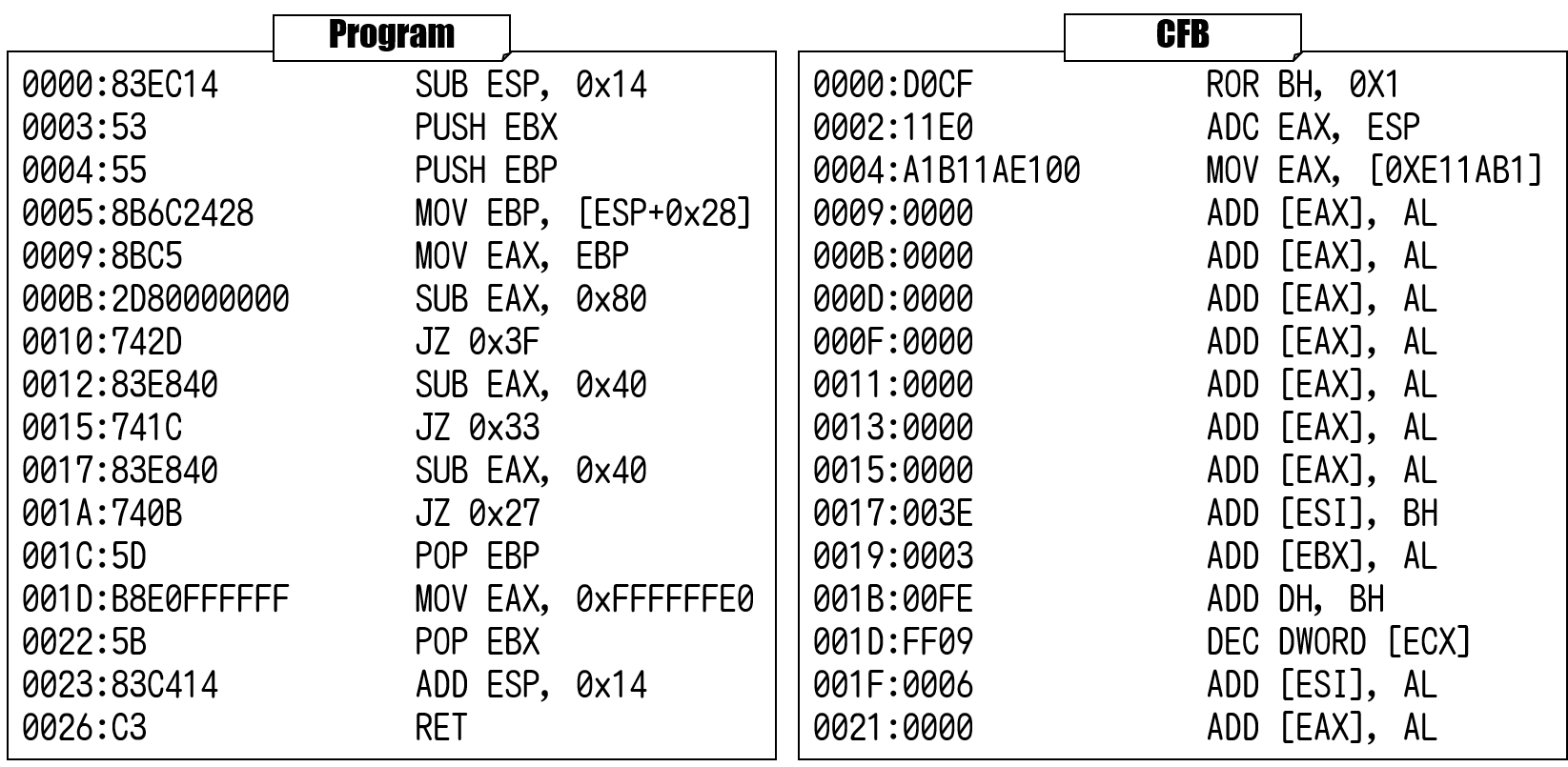}
\caption{The results of disassembling a native-code or CFB file}
\label{fig:disasm}
\end{center}
\end{figure}

\begin{figure}[tb]
\begin{center}
  \includegraphics[width=8cm,clip]{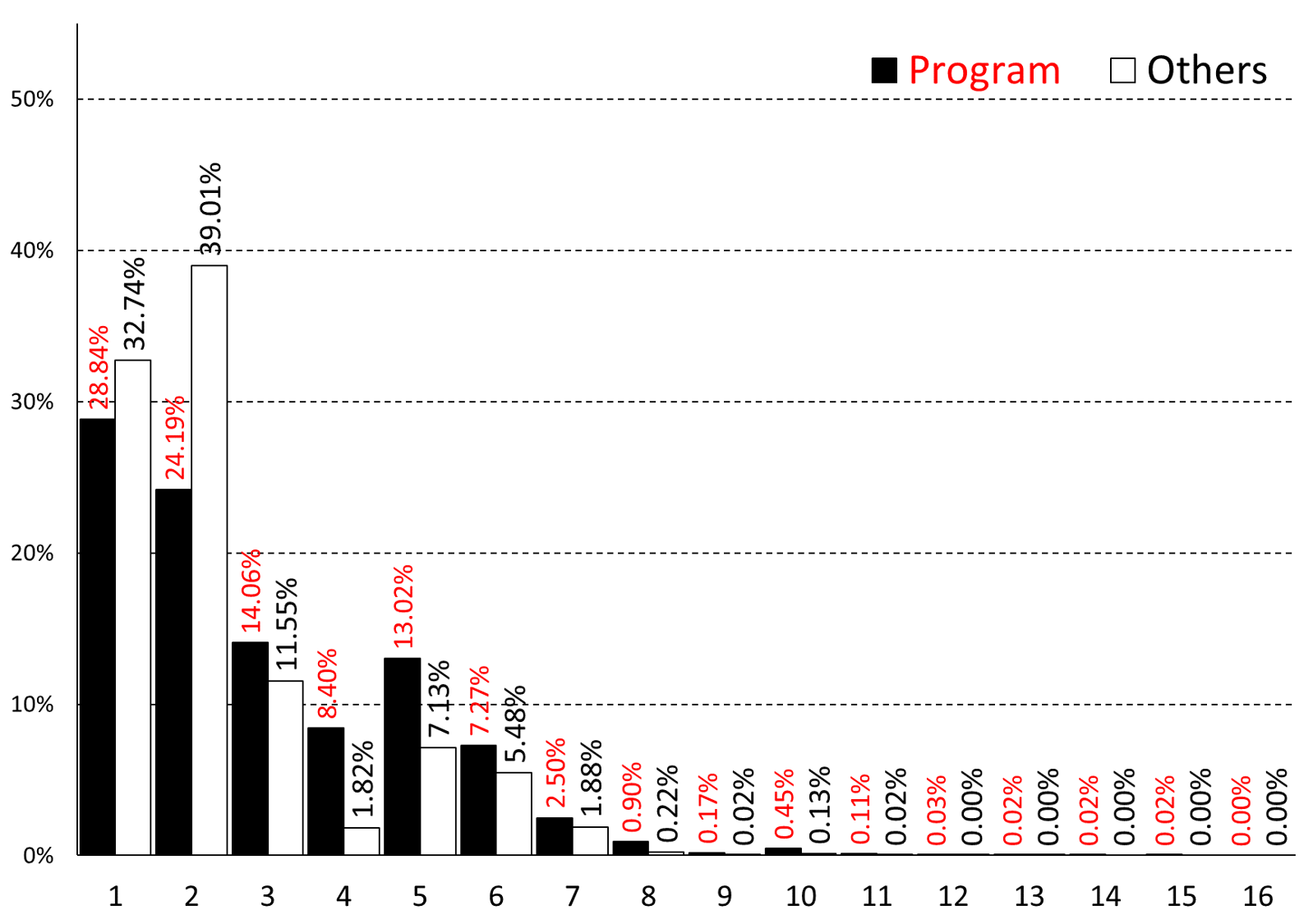}
\caption{
Distributions of instruction lengths in files from the ``Program'' and ``Others'' categories.
The total number of instructions in each category is over 600,000.
}
\label{fig:len_opcode}
\end{center}
\end{figure}

The following procedure is used to make ``Code'' datasets.
First, we treat the files of a ``File'' dataset as x86 code files, whether they come from the ``Program'' category or the ``Others '' category.
Second, we separate these files into ``instructions'' (i.e., disassemble the real or pretended x86 code).
Third, we convert each instruction into a 128-bit fixed-length instruction by padding it with ``0x00.''
Finally, packing 16 randomly selected fixed-length instructions into one set, we make a 2048-bit sequence.
The reason we padded instructions to 128 bits (16 bytes) is the following.
According to the specification of the x86 architecture~\cite{IntelManual} 15 bytes is basically the maximum length of one instruction.
Thus we padded each instruction with null bytes to achieve a fixed length of 16 bytes (one byte larger than the maximum instruction length) and combined 16 of these padded instructions to form a code segment that has a convenient length for our analysis.
Although 15 bytes is the basic maximum length of instruction, longer instructions could appear in theory (particularly when the file being interpreted as x86 code is actually a document file).
\footnote{The following sentence appears in the specification.
\begin{quote}
Exceeding the instruction length limit of 15 bytes (this only can occur when redundant prefixes are placed before an instruction).
\end{quote}}
However, we did not find any instruction longer than 15 bytes in our experiment.
The results of disassembling x86 native code and a CFB file are shown in Fig.~\ref{fig:disasm}.
The average of the lengths of each ``instruction'' is 2.95 bytes for the`` Program'' category and 2.38 bytes for the`` Others'' category.
As shown in the figure, there are various lengths of instruction in x86 CPU architecture, which appear to have no regular pattern.
The frequencies of each instruction length in files from each category are shown in Fig.~\ref{fig:len_opcode}.

\section{Proposed Visualization Methods}
\begin{figure}[tb]
\begin{center}
  \includegraphics[width=12cm,clip]{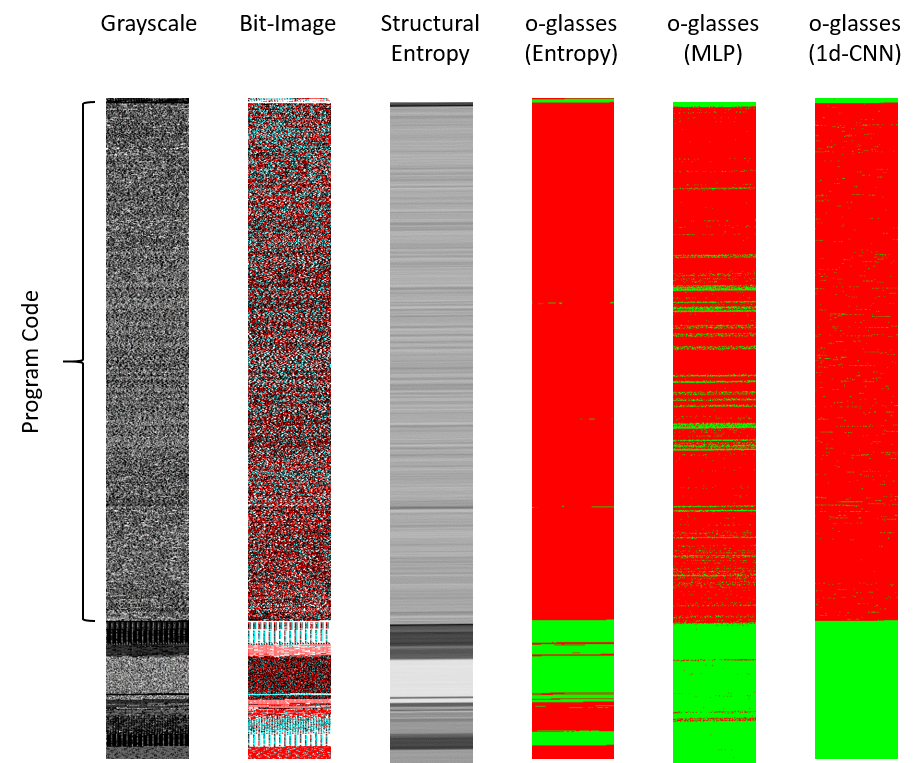}
\caption{
The result of visualizing ``notepad.exe.''
In the case of the grayscale image, we adopted the conventional conversion techniques~\cite{conti2010visual,nataraj2011malware} except for fixing a 128-pixel (byte) image width.
In the case of the structural entropy image, we selected 256 bytes for the block size.
Our methods classify the input block as either ``Program'' (red) or ``Others'' (green). 
The block sizes are 256 in o-glasses (entropy) and o-glasses (MLP), and 16 instructions in o-glasses( 1d-CNN).
The block shifts are 1 byte in all the methods.
}
\label{fig:notepad}
\end{center}
\end{figure}

In this section, we introduce three visualization methods: o-glasses (1d-CNN), o-glasses (MLP), and o-glasses (entropy), which are based on a 1d-CNN, an MLP, and entropy, respectively.
These methods classify the input block as either ``Program'' or ``Others'' and visualize the input block as an image in two colors (``Program'' is shown in red, ``Others'' is shown in green).
Fig.~\ref{fig:notepad} shows the result of visualizing ``notepad.exe'' using our methods and three conventional methods.
The details of each method are as follows.

\subsection{o-glasses (1d-CNN)}
\label{sec:1d-cnn}
First, we consider the o-glasses method based on a 1d-CNN.

\subsubsection{Local receptive field for the x86 instruction set}
We aim to make our model specialize in recognition of native program code.
If you directly input binary, such as an x86 instruction set, into a convolutional layer, you cannot identify single instructions as expected.
The input data consist of instructions serialized as one-dimensional data.
We convert the input data into N-bit fixed-length instructions to obtain features of the instructions.
Additionally, the kernel field size and the stride should be adjusted to $N$.
We selected 128 (16 bytes) as the value of $N$ because this is a convenient size that is just larger than the maximum size (15 bytes) of an x86 instruction.
In the 1d-CNN, the first layer consists of local receptive fields against each instruction.
Therefore, it is expected that the next layer obtains the relationships among instructions.

\subsubsection{Our 1d-CNN}
\begin{figure}[tb]
\begin{center}
  \includegraphics[width=12cm,clip]{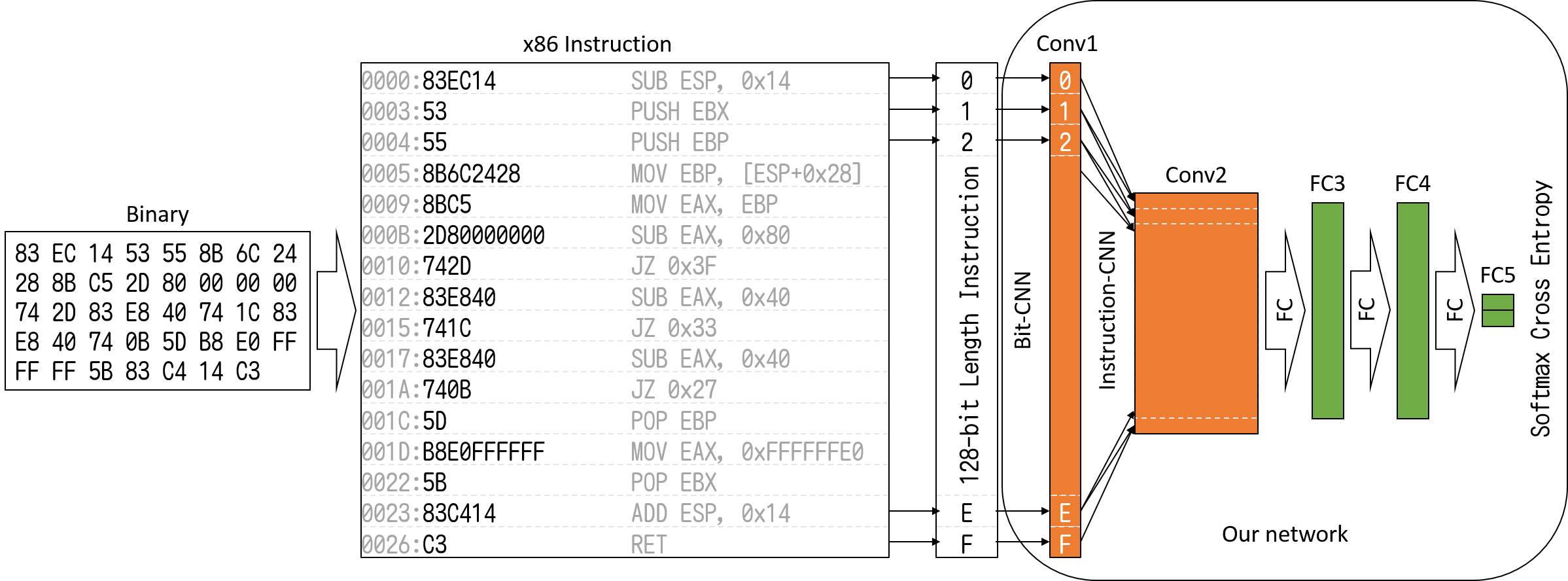}
\caption{Outline of our 1d-CNN}
\label{fig:network}
\end{center}
\end{figure}

The whole of our 1d-CNN is shown in Fig.~\ref{fig:network}.
We serialize a set of 16 fixed-length instructions into an array of 2048 bit values as input data.
The first layer is a convolutional layer (Bit-CNN).
We apply 96 layers of 128 bit-filters to a 2048 input volume.
Choosing a stride of 128, the output volume is $16 \times 96$.
The second layer is also a convolutional layer (Instruction-CNN).
We apply 256 2-filters to a $16 \times 96$ input volume with a stride of 1.
We expect that the second layer will obtain the features of the relationship between two adjacent instructions.
Our 1d-CNN does not contain any Pooling layer.
The 3rd to the 5th layers are fully connected.
Their output volumes are 400, 400, and 2, respectively.
We add two batch normalization~\cite{DBLP:journals/corr/IoffeS15} layers before the 1st and 2nd fully connected layers to speed up and stabilize the learning process.
After each layer except the last one, we apply a ReLU~\cite{glorot2011deep} layer.
The ReLU layer applies the function $f(u) = \max(u, 0)$ to all of the values in the input volume.
The softmax function is used in the final layer of our network.
\begin{equation}
{y_{k} \equiv z_{k}^{(L)} = \frac{\exp(u_{k}^{(L)})}{\sum_{j=1}^{K}\exp(u_{j}^{(L)})}
},
\end{equation}
where $K=2$ and $k \in \{1,2\}$.
Our network is trained under a cross-entropy regime.
The cross-entropy function for one training sample ($x_n$,$t_n$) for $n\in[1,N]$ is 
\begin{eqnarray}
\label{eq:error}
\begin{split}
E_{n}(\mathbf{w}) &= - \sum_{k=1}^{K} \left( t_{nk}\log y_{nk}(x_{n},\mathbf{W}) \right. \\
& \left. +(1-t_{nk})\log(1-y_{nk}(x_{n},\mathbf{W})) \right),
\end{split}
\end{eqnarray}
where the input data is $x_n \in \{0,1\}^{2048}$, the true label is $t_{n} \in \{0,1\}^{K}$, and the number of output units is $K$.
The sum of the errors $E_{n}$ calculated from each training sample is the total error function $E$:
\begin{equation}
{E(\mathbf{w}) = \sum_{n=1}^{N}E_{n}(\mathbf{w})
},
\end{equation}
where the number of samples is $N$.

\subsubsection{Stochastic gradient descent}
We use the stochastic gradient descent (SGD) method to minimize the error function in the backpropagation algorithm.
To economize on the computational cost of each iteration, SGD samples a subset of summand functions at every step. This is very effective in the case of large-scale machine learning problems.

The current weight $\mathbf{w}^{t}$ is updated to $\mathbf{w}^{t + 1}$ using the following equation.
\begin{equation}
{\mathbf{w}^{t + 1} \gets \mathbf{w}^{t} - \eta \left. \frac{\partial E(\mathbf{w})}{\partial \mathbf{w}}\right|_{\mathbf{w}=\mathbf{w}^{t}}
},
\end{equation}
where $\eta$ is the learning rate.

A compromise between computing the true gradient and the gradient of a single example is to compute the gradient against more than one training example (called a ``mini-batch'') at each step.
\begin{equation}
{E_{m}(\mathbf{w}) = \frac{1}{|N^{m}|}\sum_{n \in N^{m}}E_{n}(\mathbf{w})
},
\end{equation}
where $N^{m}$ is a subset of the index set $\{1,\ldots,N\}$ such that $\bigsqcup_{m}N^{m}=\{1,\ldots,N\}$ and $N^{m_i}\cap N^{m_j} = \O$ for $i \neq j$.

\subsection{o-glasses (MLP)}
Like the previous method, this method focuses on each block of the target file.
The input data size is one block (a 256-byte sequence), and the block shift is 1 byte.
The network containing hidden layers and the output layer is the same as fully connected layers 3--5, shown in Fig.~\ref{fig:network}, for the 1d-CNN (see Section \ref{sec:1d-cnn}).

\subsection{o-glasses (Entropy)}
The method detects program code based on whether the entropy of the block lies within a given range.
When appropriate range criteria are selected, this method achieves reasonable accuracy in the detection of program code.

\section{Evaluation}
\subsection{Recognition Performance}
We investigated the detection rates of program code by our methods using the training datasets described in Section~\ref{sec:dataset}.
Table~\ref{tab:ours} shows an overview of the results.

\begin{table}[tb] 
\caption{Performance of our methods to detect x86 code} 
\label{tab:ours}
\hbox to\hsize{\hfil
\begin{tabular}{p{9em}|p{5em}p{5em}p{5em}|p{9em}}
\hline
\hline
\hfil Our methods \hfil& \hfil F-measure \hfil& \hfil Precision \hfil & \hfil Recall \hfil & \\
\hline
Based on Entropy &\hfil 0.9266 \hfil&\hfil 0.8725 \hfil & \hfil 0.9879 \hfil&Range:0.408--0.753\\
Based on MLP&\hfil 0.9840 \hfil& \hfil 0.9830 \hfil& \hfil 0.9851 \hfil &\\
Based on 1d-CNN& \hfil 0.9995 \hfil& \hfil 0.9999 \hfil& \hfil 0.9992 \hfil&\\
\hline
\end{tabular}\hfil}
\end{table}

In the comparison of the different algorithms, we use the F-measure defined by 
\begin{equation}
{F_{\mathit{measure}} = \frac{2 \cdot \mathit{Precision} \cdot \mathit{Recall}}{\mathit{Precision} + \mathit{Recall}}
}.
\end{equation}
In this calculation, precision is given by 
\begin{equation}
{\mathit{Precision} = \frac{\mathit{TP}}{\mathit{TP} + \mathit{FP}}
},
\end{equation}
and recall is given by 
\begin{equation}
{\mathit{Recall} = \frac{\mathit{TP}}{\mathit{TP} + \mathit{FN}}
},
\end{equation}
where TP is the true positive rate, FP is the false positive rate, FN is the false negative rate, and TN is the true negative rate (see Table~\ref{tab:confusion_matrix}).
\begin{table}[tb] 
\caption{Confusion matrix} 
\label{tab:confusion_matrix}
\hbox to\hsize{\hfil
\begin{tabular}{p{7em}|p{5em}p{5em}}
\hline
\hline
\hfil Predicated \hfil& \multicolumn{2}{|c}{True Condition}\\
\hfil Condition \hfil& \hfil Program \hfil& \hfil Others\hfil \\
\hline
\hfil Program \hfil& \hfil TP \hfil& \hfil FP \hfil \\
\hfil Others \hfil & \hfil FN \hfil& \hfil TN \hfil \\
\hline
\end{tabular}\hfil}
\end{table}

\subsubsection{o-glasses (Entropy)}
We examined many ranges for the entropy rate-based binary classifier and chose the range that gives the maximum F-measure for the training dataset.
The F-measure for entropy in Table~\ref{tab:ours} was calculated using this ``range'' parameter (also shown in the table) against the test dataset.

\subsubsection{o-glasses (MLP) and o-glasses (1d-CNN)}
To train and test our network, 10-fold cross-validation was used.
After 200 epochs, we calculated the F-measure, the precision, and the recall of the test data.

Here is our parameter configuration:
\begin{itemize}
 \item $\text{learning rate}(\eta) = 0.001$
 \item $\text{mini-batch size} = 100$
\end{itemize}

\begin{figure}[tb]
    \begin{tabular}{cc}
      \begin{minipage}[t]{0.5\hsize}
\begin{center}
  \includegraphics[width=6cm,clip]{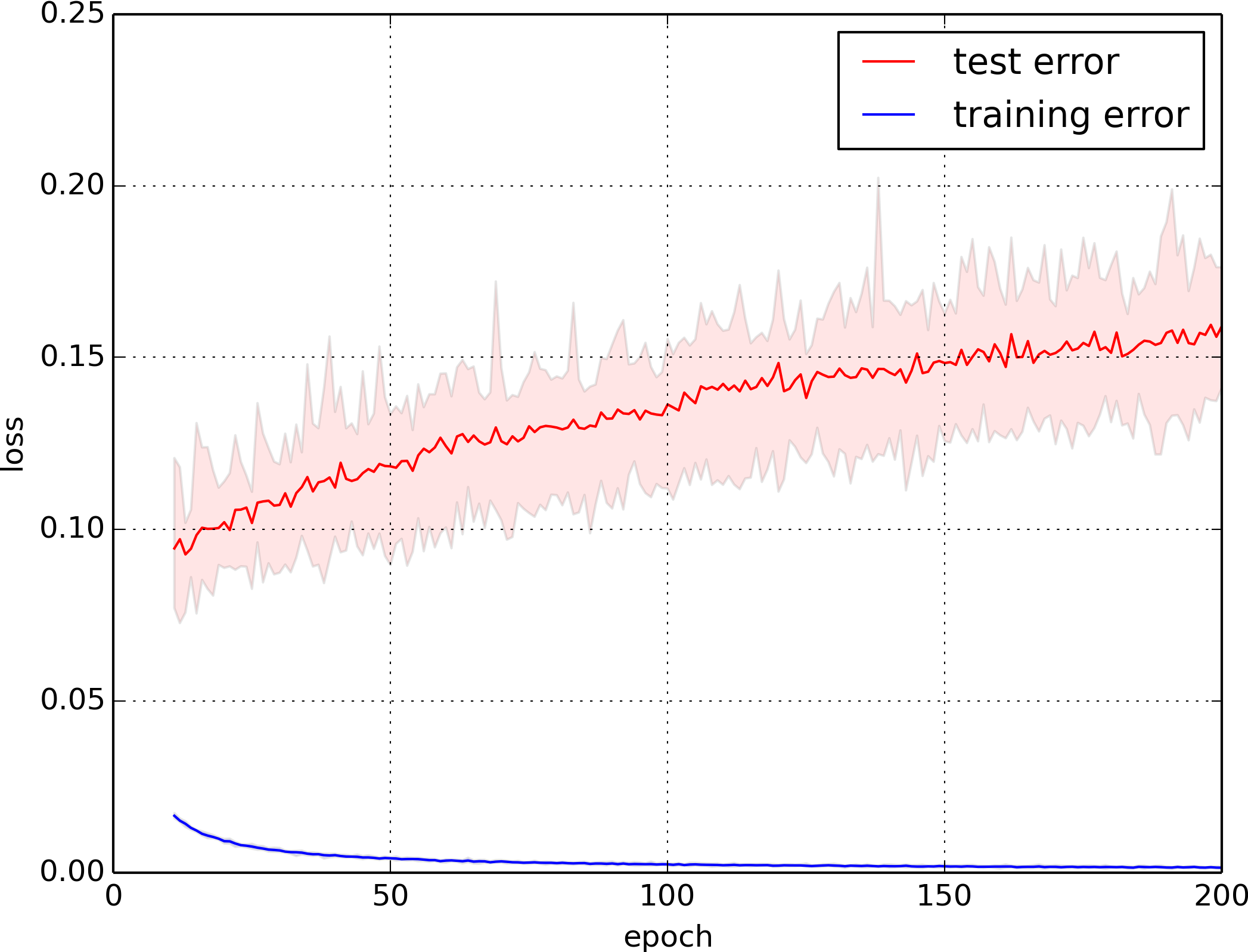}
\caption{Learning curve of o-glasses (MLP)}
\label{fig:error}
\end{center}
     \end{minipage} &
      \begin{minipage}[t]{0.5\hsize}
 \begin{center}
  \includegraphics[width=6cm,clip]{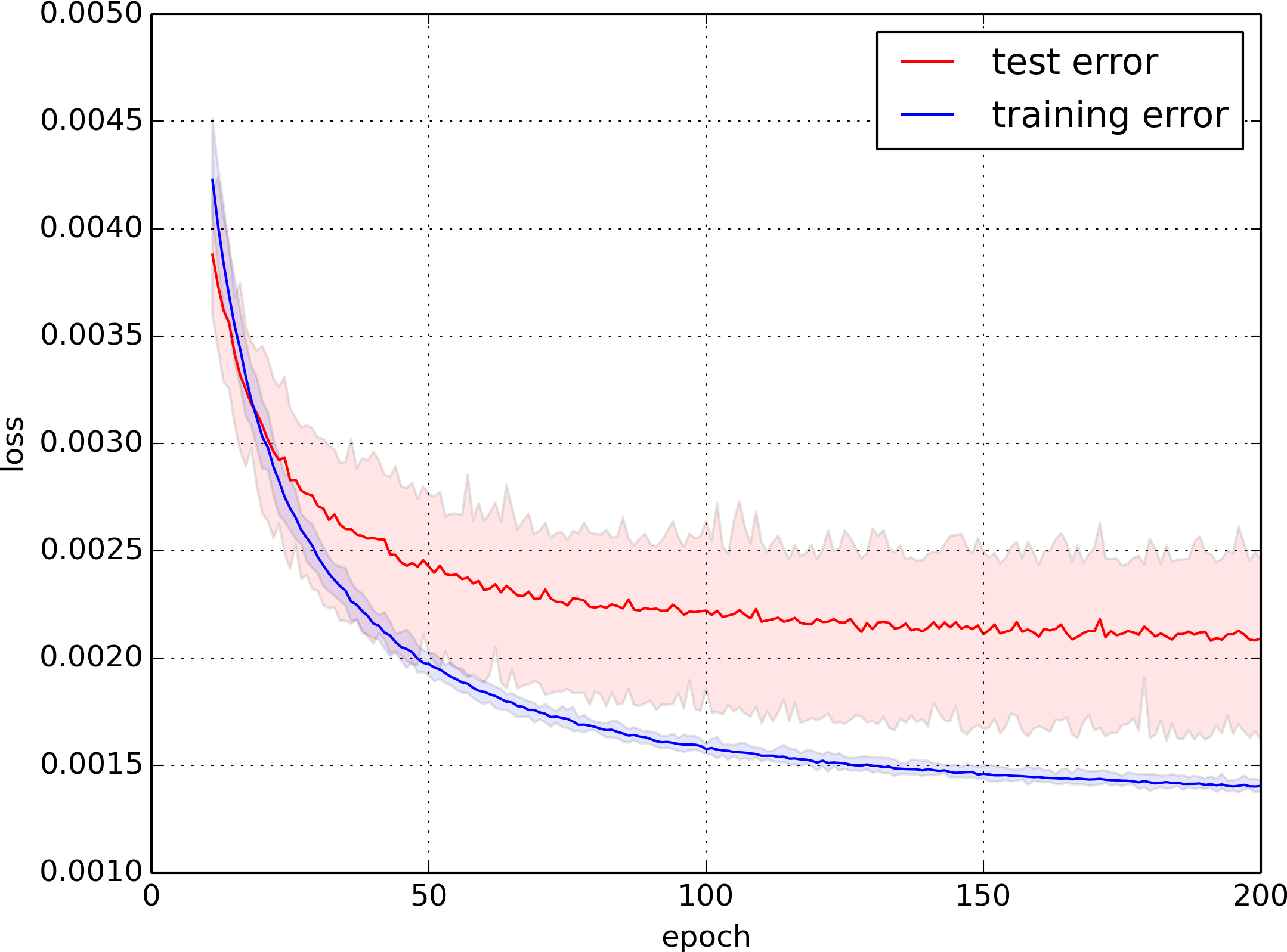}
\caption{Learning curve of o-glasses (1d-CNN)}
\label{fig:error2}
\end{center}
      \end{minipage}
    \end{tabular}
\end{figure}

The learning curves of the error are shown in Figs.~\ref{fig:error} and \ref{fig:error2}.
The blue areas indicate the range of possible values of the training errors in the 10-fold cross-validation process.
The solid blue lines indicate the average of the training errors in the 10-fold cross-validation process.
The red areas indicate the range of possible values of the test errors in the 10-fold cross-validation process.
The solid red lines indicate the average of the test errors in the 10-fold cross-validation process.
From this figure, it can be seen that our 1d-CNN method does not cause over-fitting.

\subsection{Experiments with Malicious Documents}
In this section, we visualize three malicious document files to discuss the effectiveness of our methods.
The first malicious document file contains 127 bytes of x86 code.
The second malicious document file contains 29 bytes of x86 code.
The third malicious document file does not use vulnerabilities, and does not have any x86 code.
These files are referred to in the following discussion as File 1, File 2, and File 3, respectively.

The parameters used in these experiments are the same as those described in the previous section.
After 200 epochs of training using all our datasets, we visualized the three files.

\subsubsection{File 1: CVE-2014-7247}
File 1 contains a compressed executable.
If we can analyze the file dynamically, it is easy to output the executable.
However, this file is a .jtd document file for
Ichitaro, which is Japanese word processing software similar to Microsoft Word.
The old version of Ichitaro had a vulnerability called CVE-2014-7247, which this document targets.
So, we need the old version for dynamic analysis.
When we do not have the old version, we must find the decoder for the executable file to output it.
Therefore, we need to find the shellcode.

This document file contains 127 bytes of x86 code.
The code is split two sequences; the size of the first sequence is 77bytes, and the size of the second sequence is 50 bytes.
The first sequence codes for jumping to the second sequence.

\begin{figure}[tb]
\begin{center}
  \includegraphics[width=12cm,clip]{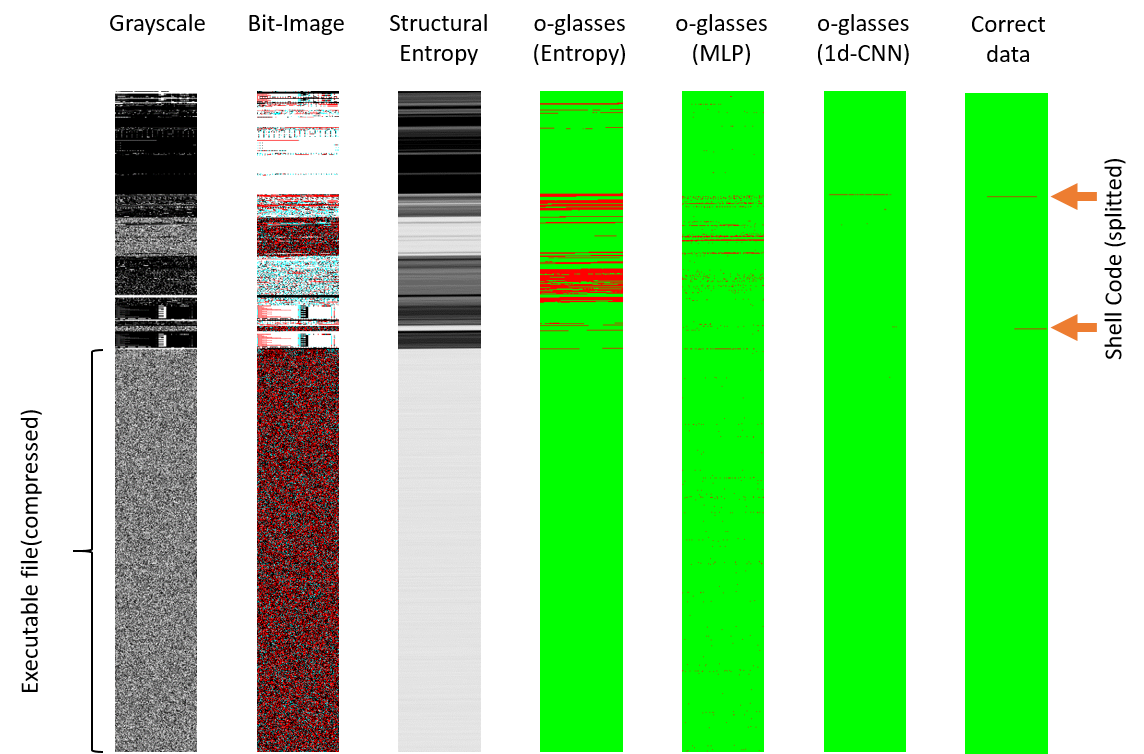}
\caption{Results of visualizing File 1 by various methods}
\label{fig:case_study1}
\end{center}
\end{figure}

Fig.~\ref{fig:case_study1} shows the result of visualizing File 1.
The o-glasses (1d-CNN) method shows an x86 code sequence at almost the same location as the first sequence.
However, the method could not locate the second sequence.

\subsubsection{File 2: CVE-2012-0158}
\begin{figure}[tb]
\begin{center}
  \includegraphics[width=12cm,clip]{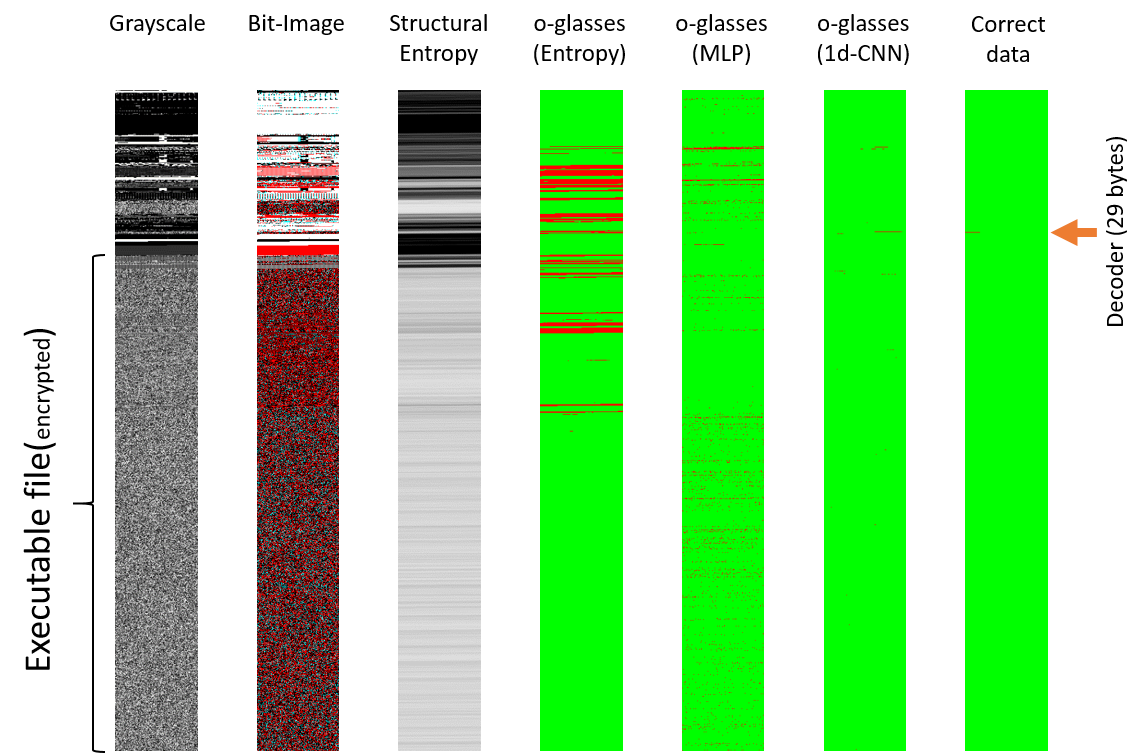}
\caption{Results of visualizing File 2 using various methods}
\label{fig:case_study2}
\end{center}
\end{figure}

File 2 contains an executable file encoded with a 2-byte-key xor.
This document file is a Word (.doc) document file and attacks a vulnerability called CVE-2012-0158.
This document file contains only 29 bytes of x86 code.
This is the code which we mentioned in the Introduction.
Fig.~\ref{fig:case_study2} shows the results of visualizing File 2.
The o-glasses (1d-CNN) method could not locate the decoder.
However, it found a sequence of ``nop'' instructions located just before the decoder.

\subsubsection{File 3: VBA script donwloader}
\begin{figure}[tb]
\begin{center}
  \includegraphics[width=9.95cm,clip]{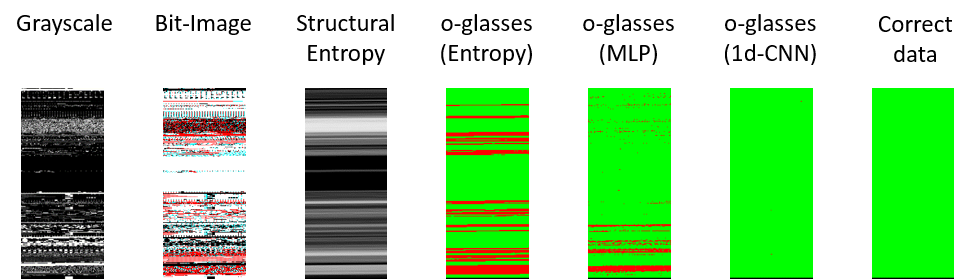}
\caption{Results of visualizing File 3 by various methods}
\label{fig:case_study3}
\end{center}
\end{figure}

Unlike the other two files, File 3 does not contain any executable file.
Additionally, this document file does not attack any vulnerabilities.
Instead, a VBA script in this document file downloads an executable file from the internet and runs it.
Therefore, this document file does not contain any x86 code.
As shown in Fig.~\ref{fig:case_study3}, o-glasses (1d-CNN) correctly reports no x86 code in this document, while the other methods report many false-positive blocks.
Thus, human examiners can confidently focus on the positive blocks reported by o-glasses (1d-CNN) to search for real shellcode in malicious documents.

\section{Discussion}
In this section, we discuss the usage and limitations of our methods, and areas for future work.

\subsubsection{Easily Collectible Training Datasets}
One of the most significant problems using machine learning is how to prepare the training dataset.
Even an excellent model cannot demonstrate its performance without large samples.
Many studies of malware using machine learning have sometimes struggled to collect samples because they need hard-to-collect malware.
In contrast, our approach does not need malware for the training dataset.
Since all we need to collect is x86 code and normal document files, it possible for anyone to create training datasets from easily accessible sources.
Surprisingly, in spite of this fact, our method, o-glasses(1d-CNN), can find the locations of shellcode almost exactly.
Therefore, our proposed methods suggest a possible beneficial effect for professional malware analysis.
On the other hand, some shellcode is known to contain garbage code.
We did not consider such cases, and therefore our dataset needs to be improved.

\subsubsection{High Recognition Rate for x86 Code}
In this paper, we have presented a method of recognizing program code in document files using a 1d-CNN.
Using a local receptive field and weight sharing, our 1d-CNN can capture important features of instructions.
Thus, even if the input instruction sequence is shifted, our network can recognize program code with a high degree of success, as measured by the F-measure rate.

The result of our experiments, inputting 16 opcodes into our network, is that the F-measure rate reaches about 99.95\%.
While this value seems to be very high at first glance, it means that,
when the target file size is 100 KB, about 50 bytes of noise is generated in the visualization result.
When looking for a small program like shellcode, this noise becomes an obstacle to analysis.
Although our method has already achieved a real-use level of performance for human analysts, it still needs further improvement for automatic shellcode detection.

\subsubsection{Visual Analysis to Support Analysts}
In this paper, we visualized several malicious document files and showed that we could find some small programs like shellcode.
Furthermore, in the case of a document file which does not contain any x86 native code, other methods do not provide convincing evidence that x86 native code was not present.
But, by using our method, we can be fairly confident that a file does not contain x86 native code.

However, some malicious document files do not contain x86 native code, but contain interpreted code such as JavaScript.
Our methods do not cover such files.
For these files, it is necessary to analyze the malware by another method, which may be combined with our o-glasses (1d-CNN) method.

\section{Conclusion}
In this paper, we proposed a 1d-CNN for detecting program code in document files.
We observed that a local receptive field for a 128-bit fixed-length instruction is effectively formed in the first layer of our network.
We can balance both high precision rate and high recall rate for detecting program code by using our network.
Our network can narrow down a target for human static analysis of unknown malware.
Future work includes increasing the number of malicious document files used to check the validity of our proposed method.
Another task is to combine our network with various analysis methods for unknown malware.

\bibliographystyle{splncs}
\bibliography{books}

\end{document}